\documentclass[aps,prl,twocolumn,preprintnumbers]{revtex4}
\usepackage{graphicx,amsmath}
\usepackage{dcolumn}
\usepackage{amsmath}
\usepackage{amssymb}
\usepackage{amsfonts}

\begin{document}

\title{Surfaces of Constant Temperature for Glauber Dynamics}         
\author{David Ford}        
\email[]{dkf0rd@netscape.net}
\affiliation{Department of Physics, Naval Post Graduate School, Monterey, California}
\date{\today}          
\begin{abstract}
The wavefunction of a single spin system in a prepared initial state evolves
to equilibrium with a heat  bath. The average spin $$q(t) = p_{\uparrow}(t) - p_{\downarrow}(t)$$ exhibits a
characteristic time for this evolution.

With the proper choice of spin flip rates, a dynamical Ising model (Glauber)
can 
be constructed with the same characteristic  time for transition of the average
spin to equilibrium.
The Glauber dynamics are
expressed as a Markoff process that possesses many of the same
physical characteristics as its quantum mechanical counterpart. 

In addition, since  the classical trajectories are those
of an ergodic process (the time averages of a single trajectory
are equivalent to the ensemble averages), the surfaces of constant temperature, 
in terms of the model parameters,  may be derived for the single spin system.
\end{abstract}
\maketitle

\section{Background on the Example}
The goal of this short note is to establish the surfaces of constant
temperature consistent with the Glauber dynamics of a simple example.
The example itself is taken from Glauber's original paper \cite{glauber}. 

The subsystem of interest is a single spin particle in equilibrium with a heat bath. 
The total system consists of particle plus bath. The physical parameters are 
applied external field strength $H$ and temperature $\theta$. The particle is 
in a prepared initial state and evolves to a mixed state in
equilibrium with the bath. 
The time constant for this evolution may be measured and used to construct
a Markoff process with classical trajectories and the same decay time.

\begin{figure}[htbp]
\begin{center}
\leavevmode
\includegraphics[width=60mm,keepaspectratio]{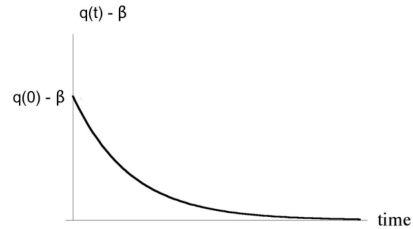}
\caption{Relaxation of known average spin to its equilibrium value $\beta$ occurs
exponentially with parameter $\alpha.$}
\label{relax}
\end{center}
\end{figure}

The inverse of the decay time is used as a rate (spin flips per unit time) parameter
in the construction of the Markoff process and is denoted by $\alpha.$ The value of the average
spin of the system at equilibrium is denoted by $\beta$ and is used to model 
the presence of a magnetic field. This situation is depicted
in figure \ref{relax}.

\begin{figure}[htbp]
\begin{center}
\leavevmode
\includegraphics[width=60mm,keepaspectratio]{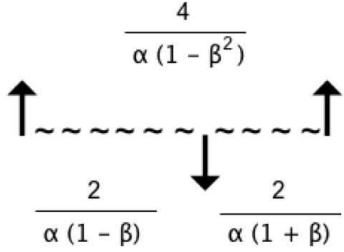}
\caption{Characteristic holding times and total Carlson depth for the states
of the single spin system}
\label{timescales}
\end{center}
\end{figure}

At equilibrium the 
probability amplitudes correspond to Gibb's distribution with Ising Hamiltonian
$(\mu H, - \mu H)$ for the up and down states respectively. Note that for all values of H 
the sum of the subsystem energies is zero \cite{ford}.  By  choosing the state transition rates
of the classical trajectories appropriately, this too may be built into the Markoff process.
See figure \ref{timescales}.

\section{Temperature Development}
A large ensemble of N identical single spin subsystems are prepared. 
Once equilibrium with the bath has been attained, measurement reveals $N_{\uparrow}$ up and $N_{\downarrow}$
down. The ratio of the state probabilities is given simply by

$$\frac{p_{\uparrow}}{p_{\downarrow}} = \frac{N_{\uparrow}}{N_{\downarrow}}.$$

If the time evolution of a Markoff process is ergodic, the state probabilities
may also be expressed in terms of the epochs of the average cycle behavior. See figure \ref{timescales}.
Per characteristic cycle time, the associated equilibrium Markoff process (as defined by the state transition
rates for the spins)  spends 

$$\Delta t_{\uparrow} = \frac{2}{\alpha ( 1 - \beta)}$$

$$\Delta t_{\downarrow} = \frac{2}{\alpha ( 1 + \beta)}$$

\noindent in the up and down states respectively. The probability ratios are given by
$$\frac{p_{\uparrow}}{p_{\downarrow}} = \frac{ \frac{2}{\alpha ( 1 - \beta)}}{\frac{2}{\alpha ( 1 + \beta)}}
=\frac{1 + \beta}{1 - \beta}.$$

At zero field, neither state $\uparrow$ or $\downarrow$ is preferred. The classical single particle system
switches from one state to the other at random. In the language of the Glauber parameters
this situation corresponds to 

$$\alpha \neq 0, \,  \beta = 0.$$

Alternatively, in terms of the system temperature and applied field

$$ \theta \neq 0, \,  H = 0.$$

Clearly, since the two languages describe the same phenomenon, there is an implied
mapping between the $\theta-$axis in $(H, \theta)-$space and the $\alpha-$axis in
the $(\alpha, \beta)-$space.  In systems whose parameterization lies purely along either the 
 $\alpha$ or  $\theta-$axes, the amount of time spent spin up per characteristic period
 is equal to the amount of time spent spin down. This implies another pair of mappings
 between these axes and the line
 $$\Delta t_{\uparrow} = \Delta t_{\downarrow}$$
 
 \noindent in the time domain.

\begin{figure}[htbp]
\begin{center}
\leavevmode
\includegraphics[width=60mm,keepaspectratio]{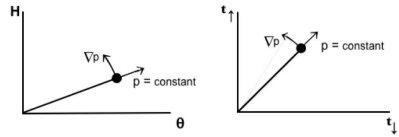}
\caption{Observers agree on lines (constant probabilities) and circular arcs 
(maximum $\nabla p$)
in both spaces.}
\label{twoPgrad}
\end{center}
\end{figure}

Note that for arbitrary constants $\lambda_1$ and $\lambda_2$, the transitions (dilatations)

$$(\Delta t_{\uparrow}, \Delta t_{\downarrow} ) \longrightarrow 
(\lambda_1 \, \Delta t_{\uparrow},  \lambda_1 \, \Delta t_{\downarrow} )$$

and

$$(H, \theta) \longrightarrow (\lambda_2 \, H, \lambda_2 \,  \theta ) $$

\noindent leave the probabilities invariant. The direction of maximum probability gradient lies perpendicular
to these invariant directions in either space. See figure \ref{twoPgrad}. 

In \cite{ford}, these simple observations are used to construct
the surfaces of constant temperature in time. That is, the image of the lines $\theta = $ constant,
in the $(H, \theta)-$space, mapped to $(\Delta t_{\uparrow},  \Delta t_{\downarrow})-$space via
the relation
$$
\theta(\Delta \mathbf{t})= \frac{ \textrm{const.} }
{ \| \Delta \mathbf{t}\|_2 \,  \sqrt{ (\log [ \frac{ \Delta t_{\uparrow} }{\Delta t_{\downarrow}} ])^2 + 1        }}.
$$

\begin{figure}[htbp]
\begin{center}
\leavevmode
\includegraphics[width=60mm,keepaspectratio]{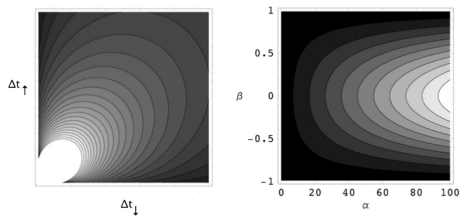}
\caption{Lines of constant temperature as seen from the time domain coordinates
($\Delta t_{\uparrow}$, $\Delta t_{\downarrow}$)  and in terms of the Markoff
parameters ($\alpha$, $\beta$).}
\label{notbad3}
\end{center}
\end{figure}

These surfaces
are presented in the left hand panel of figure \ref{notbad3}. The right hand panel of the
figure shows the same surfaces as seen from $(\alpha, \beta)$-space.

The implication is that, at constant temperature,  the  decay parameter $\alpha$
for the average spin $q(t) - \beta$
increases with increasing applied field parameter $\beta$.


\section{Bibliography}

\begin{thebibliography}{10}

\bibitem{glauber}
R. Glauber,
\newblock  `` Time Dependent Statistics of the Ising Model",
\newblock{J. Math. Physics 4, 2, pp. 294-307},
1963

\bibitem{ford}
D. Ford,
\newblock ``Surfaces of Constant Temperature in Time,''
\newblock {http://www.arxiv.org/abs/cond-mat/0510291},
2005


\end{thebibliography}
\end{document}